\documentclass{ws-ijmpa}
\usepackage{color,graphicx}
\usepackage{caption}
\begin{document}
\markboth{Y.Younesizadeh, A.A.Ahmad, A.H.Ahmed, F.Younesizadeh, M.Ebrahimkhas }{ New Black Hole Solution In Dilaton Gravity Inspired By  Power-Law Electrodynamics}

%
\catchline{}{}{}{}{}
%

\title{A New Black Hole Solution In Dilaton Gravity Inspired By  Power-Law Electrodynamics
}

\author{Younes Younesizadeh\footnote{E-mail: younesizadeh88@gmail.com}
}

\address{Department of Physics, Isfahan University of Technology, Isfahan, Iran.}

\author{Amir A.Ahmad\footnote{amir.ahmad@su.edu.krd} and Ali Hassan Ahmed\footnote{Ali.ahmed@su.edu.krd}
}

\address{Physics Department, Salahaddin University-Erbil,
 Kirkuk Road,44001 Erbil,
Kurdistan Region, Iraq.}

\author{Feyzollah Younesizadeh\footnote{fyounesizadeh@gmail.com}
}

\address{Department of physics, Faculty of science, Sahand University of Technology, Tabriz, Iran.}

\author{Morad Ebrahimkhas\footnote{Ebrahimkhas@iau-mahabad.ac.ir}
}

\address{Department of Physics, Faculty of Science, mahabad branch, Islamic Azad University, Iran, 59135-443.}

\maketitle

\begin{history}
\received{Day Month Year}
\revised{Day Month Year}
\end{history}

\begin{abstract}
 In this work, a new class of slowly rotating black hole solutions in dilaton gravity has been obtained where dilaton field is coupled with nonlinear Maxwell invariant. The background space-time is a stationary axisymmetric geometry. Here, it has been shown that the dilaton potential can be written in the form of generalized three Liouville-type potentials. In the presence of this three Liouville-type dilaton potential, the asymptotic behavior of the obtained solutions are neither flat nor (A)dS. One bizarre property of the electric field is that the electric field goes to zero when $r\rightarrow 0$ and diverges at $r\rightarrow\infty$. We show the validity of the first law of thermodynamics in thermodynamic investigations. The local and global  thermodynamical stability are investigated through the use of heat capacity and Gibbs free-energy. Also, the bounded, phase transition and the Hawking-Page phase transition points as well as the ranges of black hole stability have been shown in the corresponding diagrams. From these diagrams, we can say that the presence of the dilaton field makes the solutions to be locally stable near origin and vanishes the global stability of our solutions. In final thermodynamics analysis, we obtain the Smarr formula for our solution. We will show that the presence of dilaton field brings a new term in the Smarr formula. Also, we find that the dilaton field makes the black hole(AdS) mass to decrease for every fix values of S(entropy).

\end{abstract}

\ccode{PACS numbers:}


\section{INTRODUCTION}	
In the context of gravity the General Theory of Relativity is a successful  accomplishment. This theory parallel with quantum field theory are the basis of modern physics. General Relativity speaks with the language of differential geometry which is a different approach rather than quantum field theory and it is no any exaggeration to say that this theory actually has been revolutionized our understanding of the Universe.

The Einstein field equations which are completely unchanged since November 1915 when Einstein proposed them are still the best description of space-time behaviour in large scales. These equations are:
\begin{equation}
G_{\mu\nu}=\frac{8\pi G}{c^4}T_{\mu\nu}
\end{equation}
Where $G_{\mu\nu}$ is the Einstein tensor, $T_{\mu\nu}$ is the energy-momentum tensor, c is the speed of light and G is the Newton's gravitation constant (we set here G=c=1). It is believed that these equations govern the expansion of the universe, formation of all structures in the universe from planets, stars and solar-systems up to cluster and super-cluster galaxies, black holes behaviour, gravitational waves and so more.

The limitations of General Relativity come to existence from such scenarios like the 'dark universe'. For almost thirty years the impressive amounts of astronomical data indicate that our universe is spatially flat and experiences a phase of accelerated expansion \cite{ex0}. In the context of standard cosmology, based on General Relativity, this accelerated expansion can be explained by an unknown form of energy which is called dark energy \cite{da0}, \cite{da00}. Another way of explaining this acceleration is based on some sort of modifications for the Einstein theory of gravity. There are various kind of modifications of gravity which have been proposed in the literature since past few decades. Some of them are as Lovelock gravity \cite{lo0}, scalar-tensor theories \cite{st0} \cite{st00} \cite{ts0} \cite{ts00} \cite{h10} \cite{h11}, $f(R)$ gravity \cite{fr0}, etc. 

Among the scalar fields, dilaton field is a kind of scalar field that is coupled to the Einstein gravity in the low energy limit of string theory. In past few years this scalar field has been appeared in various theories like dimensional reduction, different models of supergravity, etc. Also, it is important to see the effect of dilaton field on the structures of the solutions. It was considered that the presence of the dilaton field can change the causal structure of space-time geometry. Although, this scalar field can change the asymptotic behaviour of the space-time as well as the thermodynamical properties of charged black hole(BH) solutions. 

Nonlinear electrodynamics has been introduced for the sake of solving some problems. This theory can eliminate the divergences in self-energy of point-like charges(The singularity of the electrical field of point-like charges) in the linear Maxwell theory. The Born - Infeld nonlinear electrodynamic actually was proposed to resolve these divergences from the linear case\cite{n1}, \cite{n2}, \cite{n3}, \cite{n4}, \cite{n5}. Also, this nonlinear model can help to remove the curvature singularities in the black hole investigations \cite{bh0}.

Another new nonlinear electrodynamics model is known as the power-law Maxwell electrodynamics. In this model, the electromagnetic  part of the Lagrangian density is $\mathcal{F}=(F_{\mu\nu}F^{\mu\nu})^p$, where p is the nonlinearity parameter. The power-law Maxwell Lagrangian is conformally invariant in higher dimensions. This kind of Lagrangian is invariant under the conformal transformation $g_{\mu\nu}\rightarrow \Omega^2 g_{\mu\nu}$ and $A_{\mu\nu}\rightarrow A_{\mu\nu}$. In the last few years, the power of Maxwell invariant(PMI) has been considered as an interesting study area in the context of geometrical physics for different dimensionality ranging \cite{d0} \cite{d1} \cite{d2} \cite{d3} \cite{d4} and more clearly in the framework of Einstein- Maxwell dilaton gravity \cite{em0} \cite{em1} \cite{em2}.

The paper in organized as the following order. In section \textcolor{red}{2}, we first introduce the Einstein-Generalized Maxwell gravity which is coupled to a dilatonic field. After that, by varying this action with respect to the metric $g_{\mu\nu}$, dilaton field $\Phi$ and the gauge field $A_{\mu}$,  we have written the equations of motion. Also, We have solved these equations in order to obtain the slowly rotating dilatonic black holes inspired by power-law electrodynamics. In the end of this section we have also showed that the dialton potential can be written as the generalized three Liouville-type potentials. In section \textcolor{red}{3}, the conserved quantities as total mass, angular momentum, electric charge and electric potential have been obtained. In section \textcolor{red}{4}, we have showed that the asymptotic behavior of the obtained solutions are neither flat nor (A)dS. Also, we have plotted the electric field diagram versus r. In this case, the electric field goes to zero when $r\rightarrow 0$ and diverges at $r\rightarrow\infty$. Section \textcolor{red}{5} is devoted to investigate the thermodynamic properties of this new slowly rotating black hole solution. The black hole temperature and entropy(S) are obtained in this section and their corresponding diagrams have been investigated as well. Through the redefinition of mass as a function of the extensive quantities Q and S and their conjugates, we have showed the validity of the first law of thermodynamics. Section \textcolor{red}{6} is dedicated to determine the local and global stability of our black hole solution. In local stability analysis due to the heat capacity calculation, we determine the type-1 and type-2 phase transition points and also the impact of the dilaton field on the black hole stability through diagrams. Although, in global stability analysis via the Gibbs free energy, the Hawking-Page phase transition is determined as well as the global stability of our solution in the presence of dilaton field. In section \textcolor{red}{7}, we study the Smarr formula. We have showed that the presence of dilaton field brings a new term in the Smarr formula. Final section as is clear is devoted to our summary and closing remarks.

\section{FIELD EQUATIONS AND THE BH SOLUTIONS}

Here we introduce the action of charged black holes in the Einstein-Generalized Maxwell gravity which is coupled to a dilaton field. The general form of this action is as \cite{a1} \cite{a2}: 
\begin{equation}
S=\frac{1}{16\pi}\int{d^4x\sqrt{-g}}\left(\mathcal{R}-2g^{\mu\nu}\nabla_{\mu}\Phi \nabla_{\nu}\Phi-V(\Phi)+\mathcal{L}(\mathcal{F},\Phi)\right)
\end{equation}

Where $\mathcal{R}$ is the Ricci scalar, $V(\Phi)$ is the potential of dilaton field. The last term in this relation is actually the representation of coupling between electrodynamics and scalar field. This term is as $\mathcal{L}(\mathcal{F},\Phi)=\big(-\mathcal{F}e^{-2\alpha\Phi}\big)^p$ where $\mathcal{F}=F_{\mu\nu}F^{\mu\nu}$(In electromagnetic context, $F_{\mu\nu}$ is defined as $F_{\mu\nu}=\partial_{\mu}A_{\nu}-\partial_{\nu}A_{\mu}$), $\alpha$ is the coupling constant and power p is the nonlinearity parameter. In special case p=1, this action reduces to the Einstein-Maxwell-dilaton gravity action. Varying this action with respect to the metric $g_{\mu\nu}$, dilaton field $\Phi$ and the gauge field $A_{\mu}$ after some simplifications, yields:
\begin{equation}
\mathcal{R}_{\mu\nu}=2\partial_{\mu}\Phi\partial_{\nu}\Phi+\frac{1}{2}g_{\mu\nu}V(\Phi)+\bigg[(p-\frac{1}{2})g_{\mu\nu}+ \frac{p}{\mathcal{F}}F_{\mu\gamma}F_{\nu}^{\gamma}\bigg]\mathcal{L}(\mathcal{F},\Phi)
\end{equation}

\begin{equation}
\nabla_{\mu}(\sqrt{-g}\mathcal{L}_{\mathcal{F}}(\mathcal{F},\Phi) F^{\mu\nu})=0
\end{equation}

\begin{equation}
\partial_{\mu}\partial^{\mu}\Phi=\frac{1}{4}\frac{\partial{V(\Phi)}}{\partial{\Phi}}+\frac{\alpha p}{2}\mathcal{L}(\mathcal{F},\Phi)
\end{equation}
In these set of equations we have $F_{\mu\gamma}F_{\nu}^{\gamma}$$\equiv$$g^{\beta\gamma}F_{\nu\beta}F_{\mu\gamma}$ and  $\mathcal{L}_{\mathcal{F}}(\mathcal{F},\Phi)\equiv\frac{\partial}{\partial\mathcal{F}}\mathcal{L}(\mathcal{F},\Phi)$.

Now, we consider the following ansatz as a 4-dimensional spherical rotating metric 
\begin{equation}
ds^2=-X(r)dt^2+\frac{dr^2}{X(r)}+2a H(r)sin^2\theta dt d\phi+f(r)^2d\Omega_{2}^2
\end{equation}

Where $d\Omega_{2}^2$ is the line element of a 2-dimensional hypersurface. In this assuming metric, X(r) H(r) and f(r) are three unknown functions of r that is our goal to find. For small rotation case, we are able to solve Eqs.(\textcolor{blue}{3})-(\textcolor{blue}{5}) to first order in angular momentum parameter a. From solving Einstein field equations, the only term which is added to non-rotating case is $g_{t\phi}$ that contains the first order of a.

Here, we consider this gauge potential:
\begin{equation}
A_{\mu}=h(r)\big(\delta_{\mu}^{t}-asin^2\theta\delta_{\mu}^{\phi}\big)
\end{equation}
Also, one can show that for the infinitesimal angular momentum 
\begin{equation}
\mathcal{F}=F_{\mu\nu}F^{\mu\nu}=-2\big(h^\prime(r)\big)^2
\end{equation}
Where h(r) is an unknown function that we try to find it(For proving this relation one can see more details in the appendix). Integration of Maxwell equation(Eq.(\textcolor{blue}{4})) gives:
\begin{equation}
F_{tr}=\frac{q e^{\frac{2\alpha p\Phi(r)}{2p-1}}}{f(r)^{\frac{2}{2p-1}}}
\end{equation}
Where q is an integration constant in this relation which is  related to the electric charge.

 By using the above ansatz(Eq.(\textcolor{blue}{6})) in Eq.(\textcolor{blue}{3}), we find the following field equations. These field equations are for tt, rr, $\theta\theta$ and $t\phi$ components,  respectively.

\begin{equation}
2f^{\prime}(r)X^{\prime}(r)+f(r)X^{\prime\prime}(r)=f(r)V(\Phi)+(3p-1)f(r)\mathcal{L}(\mathcal{F},\Phi)
\end{equation}

\begin{align*}
2f^{\prime}(r)X^{\prime}(r)+f(r)X^{\prime\prime}(r)+4f^{\prime\prime}(r)X(r)=4f(r)X(r) {\Phi^{\prime}}^2+ f(r)V(\Phi)+\\(3p-1)f(r)\mathcal{L}(\mathcal{F},\Phi) \tag{11}
\end{align*}

\begin{equation}
\bigg[X(r)(f^2(r))^\prime\bigg]^\prime-2=f^2(r)V(\Phi)+(2p-1)f^2(r)\mathcal{L}(\mathcal{F},\Phi) \tag{12}
\end{equation}

\begin{align*}
f^2(r)X(r)H^{\prime\prime}(r)+2H(r)\bigg(f(r)f^\prime(r)X^\prime(r)-1\bigg)=f^2(r)H(r)V(\Phi)+\\ \bigg((2p-1)H(r)+pX(r)\bigg)f^2(r)\mathcal{L}(\mathcal{F},\Phi) \tag{13}
\end{align*}
From Eqs. (\textcolor{blue}{10}) and (\textcolor{blue}{11}) we have
\begin{equation}
\frac{f^{\prime\prime}(r)}{f(r)}=\Phi^\prime(r)^2 \tag{14}
\end{equation}
We further assume this ansatz as well:
\begin{equation}
f(r)=\beta r^N \tag{15}
\end{equation}
Where $\beta$ and N are just two constants. By putting this ansatz in Eq.(\textcolor{blue}{14}), $\Phi(r)$ will be:
\begin{equation}
\Phi(r)=\pm \sqrt{N(N-1)}ln(r)+\Phi_{0} \tag{16}
\end{equation}
Here $\Phi_{0}$ is also an integration constant. Without the loss of generality we set $\Phi_{0}=0$. As is clear in this relation, $N$ parameter is related to the dilaton field and if we set $N=1$, dilaton field will be vanished($N\geq1$).

From Eqs. (\textcolor{blue}{7}, \textcolor{blue}{9}, \textcolor{blue}{15}, \textcolor{blue}{16}), the non-vanishing components of the electromagnetic field tensor are:
 
\begin{equation}
F_{tr}=-F_{rt} = \begin{cases}
\frac{q}{\beta^{\frac{2}{2p-1}}r^{\frac{2}{2p-1}}} & \text{for $N=1$} \\
\frac{q}{\beta^{\frac{2}{2p-1}}}r^{\frac{2N[2p(N-1)-1]}{2p-1}}& \text{otherwise} \tag{17}
\end{cases}
\end{equation}

\begin{equation}
F_{\phi r}=-{a sin\theta}F_{tr}\quad,\quad  F_{\theta \phi}=-a h(r) sin2\theta \tag{18}
\end{equation}
For h(r) in the gauge potential Eq.(\textcolor{blue}{7}), we have:
\begin{equation}
h(r) = \begin{cases}
\frac{(2p-1)q}{\beta^{\frac{2}{2p-1}}(2p-3)}r^{\frac{2p-3}{2p-1}} & \text{for $N=1$} \\
\frac{(2p-1)q}{\beta^{\frac{2}{2p-1}}[2p(2N^2-2N+1)-3]}r^{\frac{2p(2N^2-2N+1)-3}{2p-1}} & \text{otherwise} \tag{19}
\end{cases}
\end{equation}
We have assumed in this work that $\alpha$=$2\sqrt{N(N-1)}$.

By using Eqs.(\textcolor{blue}{8}, \textcolor{blue}{15}, \textcolor{blue}{16}, \textcolor{blue}{19}) in Eqs.(\textcolor{blue}{10}-\textcolor{blue}{13}), one can obtain
 
\begin{equation}
X(r)=\frac{r^{2-2N}}{\beta^2(2N-1)}-{m} r^{1-2N}-\frac{\Lambda}{N(4N-1)} r^{2N}+\Upsilon(r)  \tag{20}
\end{equation}

\begin{equation}
H(r)=-{m} r^{1-2N}-\frac{\Lambda}{N(4N-1)} r^{2N}+\Upsilon(r) \tag{21}
\end{equation}
Where
\begin{equation}
\Upsilon(r) = \begin{cases}
-\frac{2^{p}(2p-1)^2  q^{2p} r^{-\frac{2}{2p-1}}}{4\beta^{\frac{4p}{2p-1}}(2p-3)}  & \text{for $N=1$} \\
 \frac{2^{p}(2p-1)^2 p q^{2p} r^{2\frac{(2 p(N-1)^2-1)}{2p-1}}}{2\beta^{\frac{4p}{2p-1}}(2 p N^2-6pN+N+2p-1)(4pN^2-4 pN-2N+2p-1)} & \text{otherwise} \tag{22}
\end{cases}
\end{equation}
Without the loss of generality, we can set $\beta=1$ in the above equations.

One property of these solutions is that if we set $N=1$, the slowly rotating Kerr metric(Lense-Thirring metric) recovers. Also, by setting $a=0$ and $p=1$ the Reissner-Nordstrom metric recovers, too \cite{gr0}.

The curvature singularity for this solution is located at r=0(We have showed this fact by using the Ricci scalar diagrams in appendix).

If dilaton potential considers as a three Liouville-type potentials as:

\begin{equation}
V(\Phi)=2\Lambda_{1} e^{2\zeta_{1}\Phi}+2\Lambda_{2} e^{2\zeta_{2}\Phi}+2\Lambda e^{2\zeta_{3}\Phi} \tag{23}
\end{equation}
Then, we have
\begin{equation}
\Lambda_{1}=-\frac{N-1}{\beta^2(2N-1)}\quad,\quad \Lambda_{2}=-\frac{(2p-1)2^{p-1}q^{2p}(2pN^2-7pN+N+2p-1)}{\beta^{\frac{4p}{2p-1}}[(2N^2-6N+2)p+N-1]} \tag{24}
\end{equation}

\begin{equation}
\zeta_{1}=-\sqrt{\frac{N}{N-1}}\quad,\quad \zeta_{2}=\frac{2p(N-2)}{2p-1}\sqrt{\frac{N}{N-1}}\quad,\quad \zeta_{3}=\sqrt{\frac{N-1}{N}} \tag{25}
\end{equation}

In Eq.(\textcolor{blue}{23}), $\Lambda$ is a free parameter where plays the role of the cosmological constant.

\section{CONSERVED QUANTITIES}
The first conserved quantity that we introduce here in terms of the mass parameter m is the mass of black hole. As is clear, the asymptotic behaviour of the metric functions(Eqs.(\textcolor{blue}{20}) and (\textcolor{blue}{21})) is unusual. In this case, the quasilocal formalism of Brown and York must be applied for obtaining the quasilocal mass of dilaton black holes \cite{qas} \cite{adm}. Using this formalism and after some calculation we find the quasilocal mass as:
\begin{equation}
M=\frac{N \beta^2}{2}m \tag{26}
\end{equation}
Also, through the use of Komar conserved quantities \cite{j1} \cite{j2} \cite{j3}, we obtain
\begin{equation}
J=\frac{(4N-1) \beta^2}{3}m a \tag{27}
\end{equation}

Using Eq.(\textcolor{blue}{4}), the electric charge can be obtained through the modified Gauss law. This modified law can be written as:
\begin{equation}
Q=\frac{1}{4\pi}\int e^{-2\alpha p \Phi}\beta^2 r^{2N}(-\mathcal{F})^{p-1}F_{\mu\nu}n^{\mu}u^{\nu}d\Omega \tag{28}
\end{equation}
Where $n^{\mu}$ and $u^{\nu}$ are the space-like and time-like unit normals to a hypersurface of radius r, respectively. With the help of Eqs.(\textcolor{blue}{9}), (\textcolor{blue}{15}) and (\textcolor{blue}{16}) and after some calculations, the electric charge is given by:
\begin{equation}
Q=\frac{2^{p-1}q^{2p-1}}{4\pi} \tag{29}
\end{equation}

The electric potential U, measured by an observer at infinity with respect to the event horizon $r_{+}$, is defined by:
\begin{equation}
U=A_{\mu}\chi^{\mu}|_{r\rightarrow\infty}-A_{\mu}\chi^{\mu}|_{r=r_{+}} \tag{30}
\end{equation}
Here, $\chi=C \partial_{t}$ is the null generator of the event horizon and C is just a constant which may be fixed. Using Eq.(\textcolor{blue}{17}), the electric potential on the horizon will be as:
\begin{equation}
U= \begin{cases}
\frac{C q}{\beta^{\frac{2}{2p-1}}}\frac{2p-1}{2p-3}r_{+}^{\frac{2p-3}{2p-1}} & \text{for $N=1$} \\
\frac{C q}{\beta^{\frac{2}{2p-1}}}\frac{2p-1}{4pN^2-4pN-2N+2p-1}r_{+}^{\frac{4pN^2-4pN-2N+2p-1}{2p-1}}& \text{otherwise} \tag{31}
\end{cases}
\end{equation}

\section{PHYSICAL PROPERTIES OF THE SOLUTIONS}
In order to investigate the asymptotic behaviour of the obtaining solutions, we work on the $r\rightarrow\infty$ limit of X(r) function.
\begin{align*}
\lim_{r\rightarrow\infty}X(r)=-\frac{\Lambda}{N(4N-1)} r^{2N}+\qquad\qquad\qquad\qquad\qquad\qquad\qquad \\ \frac{2^{p-1}(2p-1)^2 p q^{2p} r^{2\frac{(2 p(N-1)^2-1)}{2p-1}}}{(2 p N^2-6pN+N+2p-1)(4pN^2-4 pN-2N+2p-1)}  \tag{32}
\end{align*}
In this limit as is clear, the asymptotic behaviour of the obtained solutions are neither flat nor (A)dS, but when we set $N=1$, the dilaton field will disappear and someone can find:
\begin{equation}
\lim_{r\rightarrow\infty}X(r)=1-\frac{\Lambda}{3} r^{2} \tag{33}
\end{equation}
This relation tells the asymptotic behaviour is flat($\Lambda=0$), AdS($\Lambda<0$) or dS($\Lambda>0$). As we can see, the unusual asymptotic behaviour of these solutions is a direct consequence of the dilaton field not even the non-linearity of the electrodynamics source.

Now, we want to study the behaviour of the electric field in the presence of the dilaton field. One bizarre property of the electric field is that the electric field goes to zero when $r\rightarrow 0$ and diverges at $r\rightarrow\infty$. In this case, we can say the presence of the dilaton filed reverses the behaviour of the electric field, because in the normal cases($N=1$) the electric field diverges at $r\rightarrow0$ and goes to zero when $r\rightarrow\infty$. To have a more precise understanding of this behaviour, we plot $E(r)$ versus r for different parameters. 

\begin{center}
\includegraphics[width=14cm]{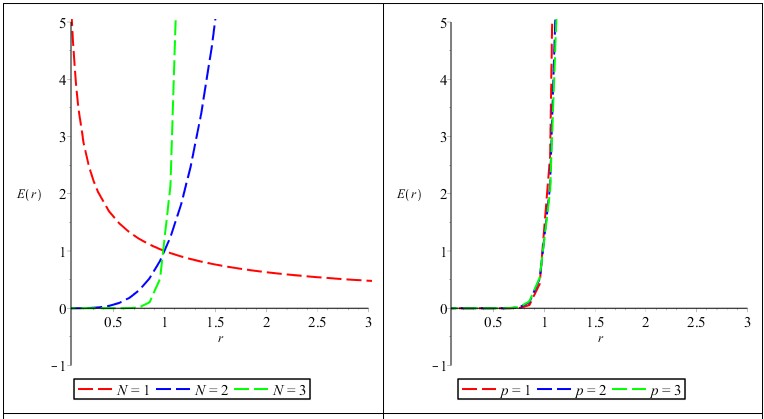}
\captionof{figure}{E versus $r$ where $N$ and p parameters change. In the left diagrams we set q=1, p=2. For the right diagrams we set  q=1, N=3.}
\end{center}

\section{THERMODYNAMICS}
From the proposing relation for the  black holes temperature via Hawking and Bekenstein, we can calculate this quantity for the previous black hole solution. The Hawking-Bekenstein temperature relation for a given  static spherically symmetric black hole is as follow through the definition of surface gravity($\kappa$):
\begin{equation}
T=\frac{\kappa}{2\pi}=\frac{1}{2\pi}\sqrt{-\frac{1}{2}(\nabla_{\mu}\chi_{\nu})(\nabla^{\mu}\chi^{\nu})}=\frac{X^{\prime}(r_{+})}{4\pi} \tag{34}
\end{equation}Where $\chi=\partial/\partial{t}$ is the killing vector of event horizon. This formula means we first take the derivative of X(r) with respect to r and then put $r_{+}$(the largest real root of the metric function $X(r_{+})=0$) instead of r. The Hawking-Bekenstein temperature for this black hole solution can be written as:

\begin{equation}
T=-\frac{(N-1)}{2\pi \beta^2(2N-1)}r_{+}^{1-2N}+\frac{(2N-1)m}{4\pi}r_{+}^{-2N}-\frac{\Lambda}{2\pi  (4N-1)}r_{+}^{2N-1}+\frac{\Upsilon^{\prime}(r_{+})}{4\pi} \tag{35}
\end{equation}
Where we have
\begin{equation}
\Upsilon^{\prime}(r_+) = \begin{cases}
\frac{2^{p}(2p-1)  q^{2p} r_{+}^{-\frac{2p+1}{2p-1}}}{2\beta^{\frac{4p}{2p-1}}(2p-3)}  & \text{for $N=1$} \\
 \frac{2^{p}(2p-1) p q^{2p}(2pN^2-4pN+2p-1) r_{+}^{\frac{4pN^2-8pN+2p-1}{2p-1}}}{\beta^{\frac{4p}{2p-1}}(2 p N^2-6pN+N+2p-1)(4pN^2-4 pN-2N+2p-1)} & \text{otherwise} \tag{36}
\end{cases}
\end{equation}
In order to see the effect of some parameters on temperature, we present the following diagrams to show the effect of $\Lambda$ and N parameters on this thermodynamical quantity. 

\begin{center}
\includegraphics[width=14cm]{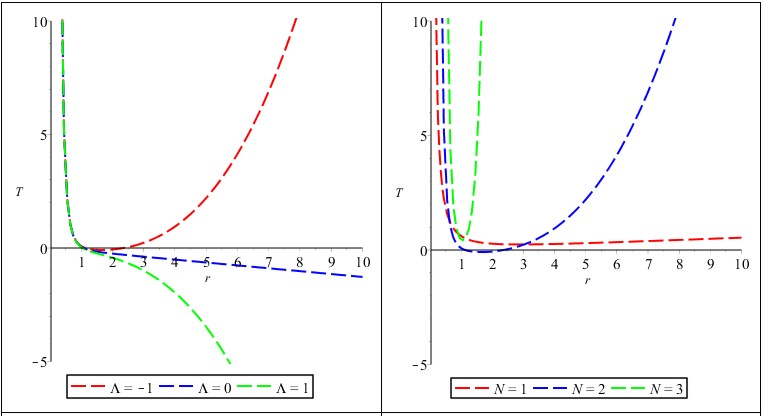}
\captionof{figure}{T versus $r_{+}$ where $\Lambda$ and k parameters change. In the left diagrams we set m=1, q=1, N=2, p=2, $\beta=1$. For the right diagrams we set m=1, q=1, $ \Lambda=-1$, p=2, $\beta=1$.}
\end{center}

The semiclassical black hole entropy(S) in n+1 dimensions from the Bekenstein-Hawking formula brings:
\begin{equation}
S=\frac{A_{+}}{4}=\frac{\omega_{n-1}}{4}(\beta  r_{+}^N)^{n-1} \tag{37}
\end{equation}
Where $\omega_{n-1}$\footnote{The surface area of a unit hypersphere in n-1 dimensions} is given by:
\begin{equation}
\omega_{n-1}=\frac{2\pi^{n/2}}{\Gamma(n/2)} \tag{38}
\end{equation}
In 3+1 dimensions we have:
\begin{equation}
S=\frac{A_{+}}{4}=\pi \beta^2 r_{+}^{2N} \tag{39}
\end{equation}

Now we introduce the mass parameter m in terms of horizon radius $r_{+}$ which is not a difficult task
\begin{equation}
m(r_{+})=\frac{1}{\beta^2(2N-1)}r_{+}-\frac{\Lambda}{N (4N-1)}r_{+}^{4N-1}+m_{k}(r_{+}) \tag{40}
\end{equation}
Here, $m_{k}(r_{+})$ is as follow:
\begin{equation}
m_{k}(r_{+}) = \begin{cases}
-\frac{2^{p-1}  q^{2p}(2p-1)^2}{2\beta^{\frac{4p}{2p-1}}(2p-3)}r_{+}^{\frac{2p-3}{2p-1}}  & \text{for $N=1$} \\
 \frac{2^{p-1}(2p-1)^2 p q^{2p} r_{+}^{\frac{4pN^2-4pN+2p-2N-1}{2p-1}}}{\beta^{\frac{4p}{2p-1}}(2pN^2-6pN+N+2p-1)(4pN^2-4p N-2N+2p-1)} & \text{otherwise} \tag{41}
\end{cases}
\end{equation}

Also, one can define the mass as a function of the extensive quantities Q and S. This function can be written as $M(S,Q)$ where the intensive quantities conjugate to S and Q are:

\begin{equation}
T=\bigg(\frac{\partial M(S,Q)}{\partial S}\bigg)_{Q} \quad,\quad U=\bigg(\frac{\partial M(S,Q)}{\partial Q}\bigg)_{S} \tag{42}
\end{equation}

Here, it is worthwhile to check the validity of the first law of thermodynamics. In differential form, we have
\begin{equation}
d M(S,Q)=\bigg(\frac{\partial M(S,Q) }{\partial S}\bigg)_{Q}dS+\bigg(\frac{\partial M(S,Q) }{\partial Q}\bigg)_{S}dQ \tag{43}
\end{equation}
After some algebraic manipulation and simplification, the above equations confirm that the first law of thermodynamics is valid, which is:
\begin{equation}
dM=TdS+UdQ \tag{44}
\end{equation}

As a matter of calculation and using Eqs.(\textcolor{blue}{31}, \textcolor{blue}{39}, \textcolor{blue}{40}, \textcolor{blue}{42}), we have 
\begin{equation}
C= \begin{cases}
  {-2\pi p} & \text{for $N=1$} \\
 \frac{4\pi N p^2}{2pN^2-6pN+N+2p-1} & \text{otherwise} \tag{45}
\end{cases}
\end{equation}

\section{LOCAL STABILITY, PHASE TRANSITION, GLOBAL STABILITY}
In this section, we study the stability(local and global) and phase transition of the slowly rotating charged black holes in dilaton gravity. First, we investigate the local stability by calculating the heat capacity and its corresponding diagrams. In this case, one can use the heat capacity relation which is

\begin{equation}
C_{Q}=T\bigg(\frac{\partial S}{\partial r_{+}}\bigg)_{Q}\bigg(\frac{\partial T}{\partial r_{+}}\bigg)_{Q}^{-1}=T\bigg(\frac{\partial S}{\partial T}\bigg)_{Q}=T\bigg(\frac{\partial^2 M}{\partial S^2}\bigg)_{Q}^{-1} \tag{46}
\end{equation}
In order to ensure that the black hole is thermally stable, the heat capacity required to be positive. In other words, this positivity confirms that the black hole is locally stable(When $C_{Q}>0$, the black hole is in a stable phase, and when $C_{Q}<0$, the black hole experiences an unstable phase). From equations (\textcolor{blue}{35}), (\textcolor{blue}{39}) and the heat capacity relation we have 
\begin{align*}
C_{Q}=\bigg(2N\pi \beta^2 r_{+}^{2N-1}\bigg)\times\qquad\qquad\qquad\qquad\qquad\qquad\qquad\qquad\qquad\qquad\qquad\qquad\\ \bigg(-\frac{(N-1)}{2\pi \beta^2(2N-1)}r_{+}^{1-2N}+\frac{(2N-1)m}{4\pi}r_{+}^{-2N}+\frac{\Lambda}{2\pi  (4N-1)}r_{+}^{2N-1}+\frac{\Upsilon^{\prime}(r_{+})}{4\pi}\bigg)\\ \times \bigg(\frac{(N-1)}{2\pi \beta^2}r_{+}^{-2N}-\frac{2N(2N-1)m}{4\pi}r_{+}^{-2N-1}+\frac{(2N-1)\Lambda}{2\pi  (4N-1)}r_{+}^{2N-2}+\frac{\Upsilon^{\prime\prime}(r_{+})}{4\pi}\bigg)^{-1} \tag{47}
\end{align*}
Where $\Upsilon(r_{+})$ is the Eq.(\textcolor{blue}{22}).

It is worthwhile to introduce two important thermodynamical points. These points are bounded point and phase transition point where:
\begin{equation}
 \begin{cases}
T=0& \text{bounded point or type one phase transition} \\
 (\frac{\partial T}{\partial r_{+}})_{Q}=0 & \text{phase transition point or type two phase transition} \tag{48}
\end{cases}
\end{equation}
In other words, the bounded point is where the heat capacity or temperature vanishes. The phase transition point is where the heat capacity diverges. Here, we show the heat capacity diagrams for different parameters. In these diagrams the phase transition points are clear to see.

\begin{center}
\includegraphics[width=14cm]{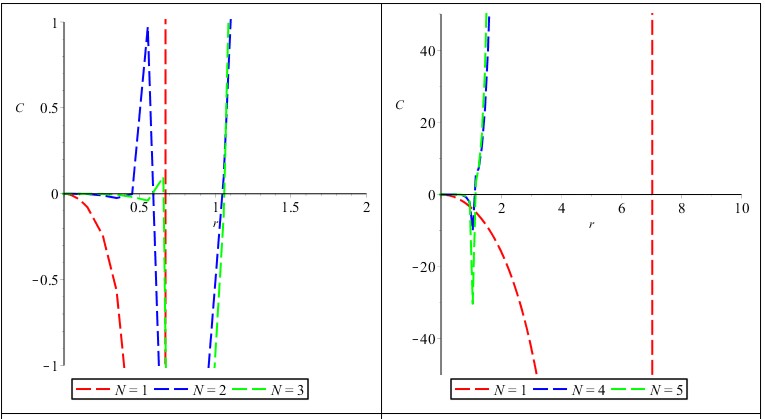}
\captionof{figure}{C versus $r_{+}$ where N changes. we set  $\Lambda=-1$, m=0.1, p=2, q=0.1 for left diagrams. $\Lambda=-0.1$, m=2, p=2, q=1 for right diagrams.}
\end{center}

\begin{center}
\includegraphics[width=14cm]{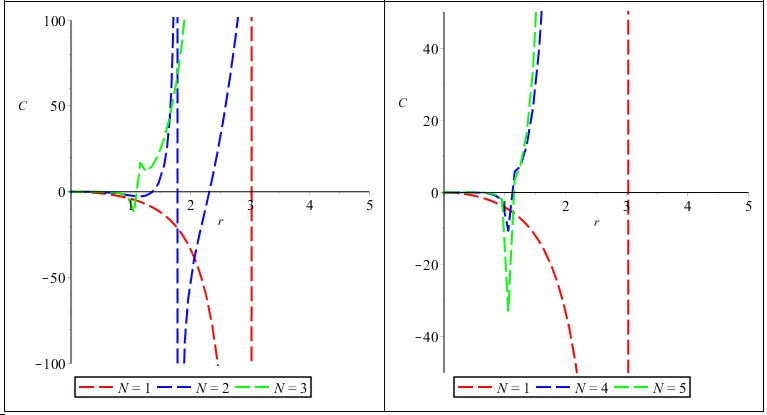}
\captionof{figure}{C versus $r_{+}$ where N changes. we set  $\Lambda=-1$, m=2, p=2, q=1 for left diagrams. $\Lambda=-1$, m=2, p=2, q=1 for right diagrams.}
\end{center}

In Fig.3, we study the effect of dilaton field on the local stability of our solutions. In the left case, we compare two cases. First, when we do not have dilaton field(N=1) , second when we have this field($N\geq2$). For the first case, it goes under a unstable phase until it enjoys a type-2 phase transition(around $r=0.7$). In second case which contains the blue and green colors, we can see the stability before $r=0.7$. In the right case, we compare two cases as well. First, when we do not have dilaton field($N=1$) and second we have dilaton field with $N=1,2$. As is clear, the red case is behaves as the left diagram case. It is unstable until it enjoys a type-2 phase transition(around $r=7$). When we have dilaton field($N=1, 2$), the blue and green color diagrams are unstable until they experience a type-1 phase transition($r=1$). After this point, they will be stable completely. From the Fig.3, we can say that the presence of the dilaton field makes the solutions to be stable near the origin in compare with the ordinary case(when we do not have dilaton field $N=1$). This investigations are true for Fig.4 as well.

Now, we investigate the global stability by considering the Gibbs free energy.
\begin{equation}
G=M-TS-QU \tag{49}
\end{equation}
In order to say that the black hole is globally stable depends on the positivity of the Gibbs free energy with positive temperature. Now, it is important to emphasis that the Hawking-Page phase transition occurs in places where the Gibbs free energy vanishes. The idea of global stability and phase transition was first proposed by  Hawking and Page \cite{HP}. 
Here, we plot the Gibbs free energy diagrams to see the effect of some parameters on this quantity, schematically.
\begin{center}
\includegraphics[width=14cm]{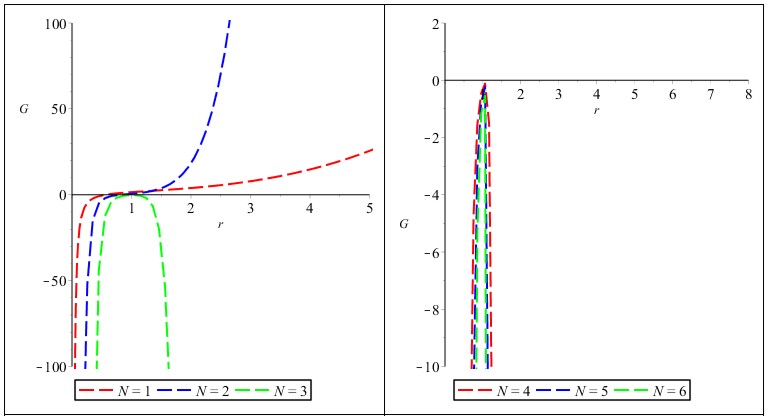}
\captionof{figure}{G versus $r_{+}$. In the left diagram $\Lambda$ changes and we set  q=1, N=2, m=1,  p=2, $\beta=1$. In the right diagram where N changes, we set $\Lambda=-1 $, q=1, m=1, p=2, $\beta=1$.}
\end{center}

In Fig.5, we want to see the effect of dilaton field on the global stability of our solutions compare to the ordinary case($N=1$). In the left diagrams, the ordinary case(N=1) is unstable until it goes under the Hawking-Page phase transition(around r=1). After that it is stable completely. When we have the dilaton field($N\geq2$), the $N=2$ case behaves like the ordinary case, but for the case of $N=3$, we encounter the complete global unstability. In the right diagrams for the greater values of N, there is no any global stability. In sum, we can say the presence of diaton field makes the solutions to be globally unstable in compare to the ordinary case($N=1$).

\section{SMARR FORMULA}
In this section, we introduce the smarr formula for our solution. As is clear, this solution leads to a new kind of relation for the smarr formula. Smarr's formula has been studied a lot in the case of non-linear electrodynamics \cite{sn0} \cite{sn1}. From Eqs.(\textcolor{blue}{26}, \textcolor{blue}{31}, \textcolor{blue}{35}, \textcolor{blue}{39}, \textcolor{blue}{45}) and after some tedious calculations, we obtain:
\begin{align*}
M=\frac{2N}{2N-1}TS+\frac{N(N-1)}{(2N-1)^2}\bigg(\frac{S}{\pi \beta^2}\bigg)^{\frac{1}{2N}}-\frac{\beta^2N\Lambda}{(2N-1)(4N-1)}\bigg(\frac{S}{\pi \beta^2}\bigg)^{\frac{4N-1}{2N}}\\ -\frac{\big(2pN^2-4pN+2p-1\big)}{ p(2N-1)}Q U \tag{50}
\end{align*}
We name this equation as the smarr formula for the dilatonic black holes inspired by power-law electrodynamics. In this relation, the second term is added to the Smarr formula due to the presence of dilaton field, because when we set $N=1$ the dilaton field will disappear(one can check the Eq.(\textcolor{blue}{16})). Now, we want to see the changes of M(mass parameter) versus S(entropy). As is clear, we find that the dilaton field makes  the M(mass parameter) to decrease for every fix values of S. This fact is obvious in the following diagrams. We have showed the fixed value(just one case for example) with a vertical yellow line. As someone can see the red line is belongs to the ordinary case(no dilaton field $N=1$) and the other colors represent the presence of the dilaton field($N=1, N=2$). In the presence of this field, the mass would be small and smaller.

We note that this statement is not always true for the temperature T, electric potential U and the electric charge.

\begin{center}
\includegraphics[width=14cm]{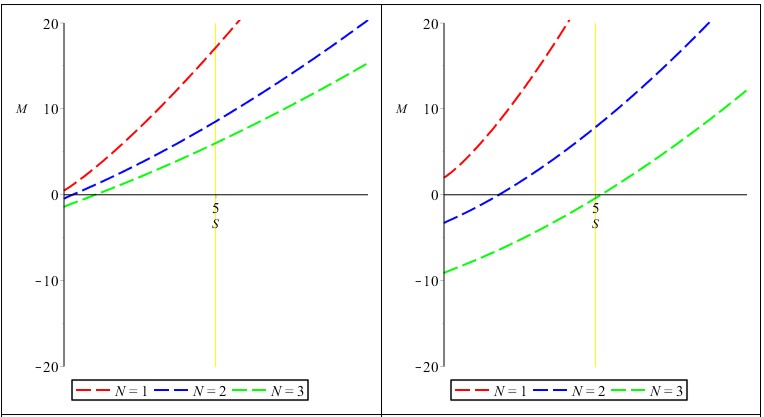}
\captionof{figure}{M versus S. we set T=1, U=1, Q=1, p=2, $\Lambda=-10$(left diagrams) when N changes. we set T=1, U=3, Q=1, p=3, $\Lambda=-20$(right diagrams) when N changes.}
\end{center}

\section{CONCLUDING REMARKS}
In this paper, we have presented a new class of slowly rotating black hole solutions in dilation gravity where dilaton field is coupled with nonlinear Maxwell invariant. From the equations of motion, we obtained the different components of field equations. We used these field equations to find an exact solution for this special case. For this solution, we showed that the dilaton field can be written as generalized three Liouville-type potentials. From different formalisms, we obtained some conserved quantities like the total mass, angular momentum, electric charge and the electric potential. In thermodynamic analysis, we have obtained the temperature and entropy(S) relations from geometrical methods. The validity of the first law of thermodynamics is also proved in these analysis. Also, we introduced the heat capacity and the Gibbs free energy to study the local and global stability of our solution, respectively. After some tedious calculation, we have obtained the Smarr formula as well. Due to the presence of dilaton field, a new term has been added to the smarr formula. We also showed that the dilaton field makes the black hole(AdS) mass to decrease for every fix values of S(entropy).

\section{APPENDIX}
\subsection{1}
The Ricci scalar relation for our solution is as follow:
\begin{equation}
\mathcal{R}=\frac{f^2(r)X^{\prime\prime}(r)+4f(r)X(r)f^{\prime\prime}(r)+4f(r)f^{\prime}(r)X^{\prime}(r)+2X(r)(f^{\prime}(r))^2-2}{f^2(r)} \tag{51}
\end{equation}
We have plotted the Ricci scalar for different parameters. In these diagrams this geometrical parameter diverges at r=0
\begin{center}
\includegraphics[width=14cm]{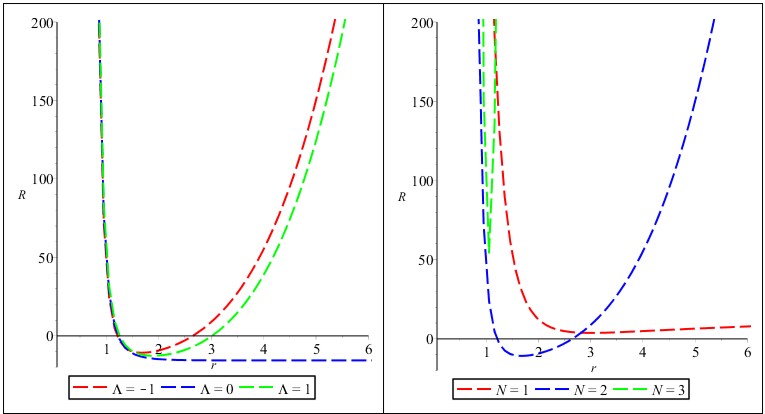}
\captionof{figure}{$\mathcal{R}$ versus r. we set m=1, q=1, N=2, p=2, $\beta=1$ when $\Lambda$ changes and m=1, q=1, $ \Lambda=-1$, p=2, $\beta=1$ when N changes. }
\end{center}
\subsection{2}
Here, we want to prove the following relation for our proposing gauge potential Eq.(\textcolor{blue}{7}) and for the slowly rotating case
\begin{equation}
\mathcal{F}=F_{\mu\nu}F^{\mu\nu}=-2\big(h^\prime(r)\big)^2 \tag{52}
\end{equation}
The non-vanishing components of this relation are:
\begin{equation}
\mathcal{F}=F_{\mu\nu}F^{\mu\nu}=F_{tr}F^{tr}+F_{rt}F^{rt}+F_{r\phi}F^{r\phi}+F_{\phi r}F^{\phi r}+F_{\theta \phi}F^{\theta \phi}+F_{\phi \theta}F^{\phi \theta} \tag{53}
\end{equation}
From the anti-symmetric property $F_{\mu\nu}=-F_{\nu\mu}$, $F^{\mu\nu}=-F^{\nu\mu}$ we have
\begin{equation}
\mathcal{F}=F_{\mu\nu}F^{\mu\nu}=2\bigg(F_{tr}F^{tr}+F_{r\phi}F^{r\phi}+F_{\theta \phi}F^{\theta \phi}\bigg) \tag{54}
\end{equation}
Also by using $F^{\mu\nu}=g^{\mu \alpha} g^{\nu \beta} F_{\alpha \beta}$, one can write

\begin{align*}
\mathcal{F}=F_{\mu\nu}F^{\mu\nu}=\qquad\qquad\qquad\qquad\qquad\qquad\qquad\qquad\qquad\qquad\\2\bigg(g^{tt} g^{rr}( F_{tr})^2+g^{t\phi} g^{rr} F_{tr}F_{\phi r}+g^{rr} g^{\phi t} F_{r\phi}F_{rt}+g^{rr} g^{\phi \phi}( F_{r\phi})^2+g^{\theta\theta} g^{\phi \phi}( F_{\theta \phi})^2\bigg) \tag{55}
\end{align*}
Using the metric ansatz Eq.(\textcolor{blue}{6}) we obtain
\begin{align*}
\mathcal{F}=F_{\mu\nu}F^{\mu\nu}=\qquad\qquad\qquad\qquad\qquad\qquad\qquad\qquad\qquad\qquad\\2\bigg(-(h^{\prime}(r))^2+\frac{2a^2H(r)}{f^2(r)}(h^{\prime}(r))^2sin^2\theta-\frac{a^2}{X(r)f^2(r)}(h^{\prime}(r))^2sin^2\theta+\frac{4a^2}{f^4(r)}h^2(r)cos^2\theta\bigg) \tag{56}
\end{align*}
For the slowly rotating case we assume $a^2=0$(we drop the second order or higher orders of a), then we arrive:
\begin{equation}
\mathcal{F}=F_{\mu\nu}F^{\mu\nu}=-2\big(h^\prime(r)\big)^2 \tag{57}
\end{equation}


\end{document}